\documentclass[12pt,english,dvips]{scrartcl}
\usepackage{babel}
\usepackage[T1]{fontenc}
\usepackage[ansinew]{inputenc}

\begin{document}

\begin{titlepage} \vspace{0.2in} 

\begin{center} {\LARGE \bf 
Canonical Quantization of Gravity \\ 
without ``Frozen Formalism'' 
\\} \vspace*{0.8cm}
{\bf Giovanni Montani
}\\ \vspace*{1cm}
ICRA---International Center for Relativistic Astrophysics\\ 
Dipartimento di Fisica (G9),\\ 
Universit\`a  di Roma, ``La Sapienza",\\ 
Piazzale Aldo Moro 5, 00185 Rome, Italy.\\ 
e-mail: montani@icra.it\\ 
\vspace*{1.8cm}

PACS 83C  \vspace*{1cm} \\ 

{\bf   Abstract  \\ } \end{center} \indent
We write down a quantum gravity equation which generalizes 
the Wheeler-DeWitt one in view of including 
a time dependence in the wave functional. 
The obtained equation provides a consistent 
canonical quantization of the 3-geometries 
resulting from a ``gauge-fixing'' (3 + 1)-slicing of the 
space-time.

Our leading idea relies on a criticism to the possibility 
that, in a quantum space-time, the notion of a (3 + 1)-slicing 
formalism  (underlying the Wheeler-DeWitt approach) 
has yet a precise physical meaning.\\ 
As solution to this problem we propose of adding to the 
gravity-matter  
action the so-called {\em kinematical action}  
(indeed in its reduced form, as implemented in the quantum regime), 
and then we impose the new quantum constraints. 

As consequence of this revised approach, 
the quantization procedure of the 3-geometries 
takes place in a fixed reference frame and 
the wave functional acquires     
a time evolution along a one-parameter family of spatial 
hypersurfaces filling the space-time.\\ 
We show how the states of the new quantum 
dynamics can be arranged into 
an Hilbert space, whose associated inner product induces a 
conserved probability notion for the 3-geometries. 

Finally, since the constraints we quantize violate  
the classical symmetries 
(i. e. the vanishing nature of
the super-Hamiltonian),
then a key result is to find 
a (non-physical) restriction on the initial 
wave functional phase, 
ensuring that general relativity outcomes when taking the 
appropriate classical limit. 
However we propose a physical interpretation
of the kinematical variables which, based
on the analogy with the so-called
{\em Gaussian reference fluid}, 
makes allowance even for such
classical symmetry violation.

\end{titlepage}

\section{General Statements}

It is commonly believed that the canonical methods 
of quantization, in spite of their successful predictions 
on many elementary particles phenomena (it stands up the 
great agreement of quantum electrodynamics with experimental data), 
nevertheless they can not be extrapolated up to arbitrarily 
high energies (i. e. arbitrarily small distances), 
when the ``granular'' nature of the fields and their 
interaction with the underlying space-time are expected to be 
important.\\ 
Indeed this point of view is confirmed by the large number of 
renormalization procedure are required to get satisfactory 
predictions already  on Minkowski space 
and moreover by the inconsistencies or ambiguities 
characterizing the canonical quantization 
of ``matter'' field 
when referred to a curved space-time \cite{BD82}.

The implementation of the canonical formalism 
to the gravitational field quantization, leads to the so-called 
Wheeler-DeWitt equation (WDE) \cite{D67,MTW73}, consisting of 
a functional approach in which the states of the theory are 
represented by wave functionals taken on the 3-geometries and, 
in view of the requirement of general covariance, they do not 
possess any real time dependence.\\ 
Due to its hyperbolic nature, the WDE is characterized 
by a large number of 
unsatisfactory features \cite{D97} 
which strongly support the idea that is 
impossible  any straightforward extension to the gravitational 
phenomena of procedures well-tested only in limited 
ranges of energies; 
however in some contexts, like the very early cosmology 
\cite{H88,KT90} (where a suitable internal time variable 
is provided by the universe volume) the WDE is not a dummy theory 
and give interesting information about the origin of our 
classical universe, see \cite{KM97}, which may be 
expected to remain qualitatively valid even for 
the outcoming of a more consistent approach. 

Over the last ten years the canonical quantum gravity found its 
best improvement in a reformulation of the constraints problem in 
terms of the so-called {\em Ashtekar variables}, 
leading to the {\em loop quantum gravity theory} \cite{Ro97,QG92}; 
this more recent approach overcomes some of the WDE limits, like 
the problem of constructing an appropriate Hilbert space, 
but under many aspects is yet a theory in progress.\\ 

The aim of this paper, more than to be the answer to
the hard question whether canonical quantum gravity has some 
predictive issue, concerns a criticism on a fundamental 
{\em ansatz} on which is based the whole formalism:
the {\em ansatz} we criticize is that, in a fully 
covariant quantum theory has 
yet meaning to speak of an Arnowitt-Deser-Misner (ADM) 
formalism \cite{ADM62,MTW73} to perform a (3 + 1)-slicing of the 
space-time (as implicitly assumed in the WDE approach).\\ 
Indeed, the space or time character of a vector, 
in particular of the normal to a 3-hypersurface, is sensitive to 
the metric field and can not be definite precisely when 
the gravitational field is in a quantum state 
(unless we consider simply a perturbation theory).\\ 
By other words we claim that the quantization procedure does not 
commutes with the space-time slicing operation; 
therefore only two approaches appear to be 
acceptable and self-consistent 
(see also the beginning of Section 4):\\ 
i) The requirement of a covariant quantization of gravity 
is preserved, but any slicing of the space-time prevented,\\ 
ii) The slicing representation of the space-time is allowed, 
but any notion of full covariance given up.\\ 
The first point of view was addressed in \cite{HH83} 
and find a promising development in the recent issue of 
the so-called {\em spin foam} formalism \cite{O01}.\\ 
The analysis here presented addresses the 
second point of view as leading 
statement and by using the discussion of the canonical methods 
of quantization presented in \cite{K81}, 
as a fundamental paradigm, we get 
a reformulation of the WDE which 
overcomes many of its shortcomings. 

In \cite{K81} is provided a wide discussion about the canonical 
methods of quantization as referred to different systems 
(with a satisfactory guide to the previous 
literature on this subject) 
in order to construct a quantum geometrodynamics on the base 
of its similarities and differences 
with other theories 
(for a valuable and more recent review on the canonical 
quantum gravity and the related problem of a physical time 
see \cite{I92}. 
Of particular interest is to be regarded for our purposes
the analysis concerning the quantum theory of 
``matter'' field on a fixed background, which shows the 
necessity of adding the so-called {\em kinematical action} 
to achieve a satisfactory structure for the quantum constraints. 

Our main statement is that the kinematical term may 
be retained even in the gravitational case 
since in the ADM action 
the lapse function and the shift vector 
can be thought as assigned 
(like in the ``matter'' field case) up to a restriction 
on the initial Cauchy problem, i. e.  
the super-Hamiltonian and super-momentum constraints 
have to be satisfied on the initial 
space-like hypersurface; indeed to fix the reference frame, 
i. e. the lapse 
function and the shift vector, is equivalent to loss the 
super-Hamiltonian and the super-momentum constraints which
(equivalent to the $0-\mu$-components of the 
Einstein equations), 
however, have an envolutive character, i. e. 
if satisfied on the initial hypersurface, 
they are preserved by the dynamical evolution \cite{K81}.\\ 
Thus from a classical point of view, 
we add the kinematical action to 
the gravity-matter one, 
but we preserve the general relativity framework simply by 
assigning a particular Cauchy problem.
When passing to the quantization procedure we get equations 
describing the evolution of a wave functional 
no longer invariant under time displacements
(the invariance under 3-diffeomorphisms is yet retained), 
but therefore 
characterized by a dependence on the choice of the spatial 
hypersurface, i. e. the ``time variable''. 

By taking the evolution of the wave functional along a 
one-parameter family of spatial hypersurfaces, 
filling the space-time, 
we show how the space of the solutions for the wave equation 
can be  turned into an Hilbert space and the quantum dynamics 
can be reduced to a Schr\"odinger-like approach with 
an associated eigenvalue problem; 
it is important to stress that the quantum evolution 
of the 3-geometries is, at the end, expressed directly in terms 
of the parameter labeling the hypersurfaces 
(the so-called label time). 

As last but fundamental achievement we find the quantum 
correspondence of the restriction imposed 
classically on the Cauchy data to provide general relativity;
indeed we show that if the phase of the initial wave functional 
satisfies the usual Hamilton-Jacobi equation, then the classical 
limit $\hbar\rightarrow 0$ always reproduces 
the Einsteinian dynamics. This feature of the model is 
crucial for its viability because ensures that, though 
violated on quantum level, 
the right symmetries 
(i. e. the vanishing nature of the super-Hamiltonian) 
are restored on the classical limit.

Since the approach here presented overlap in the formalism 
the so-called multi-time formulation of canonical 
quantum gravity, 
as well as its smeared Schr\"odinger version,\cite{IK85,I92}, 
it is worth stressing differences
and similarities.
Other important approaches based on the
same (so-called) 
embedding variables, and even referred to
the path integral formalism, can be found
in \cite{H91a}-\cite{H91c}
(see also \cite{Tom}).\\

In this well-known quantization scheme, the kinematical
variables are identified with 
non-physical degrees of freedom of the gravitational field, 
extracted by an ADM resolution of the Hamiltonian constraints. 
Thus, 
unlike for our proposal, where the kinematical variables 
are added by hand (as for any other field), 
in the multi-time approach,  
no violation takes place of the classical symmetries.\\ 
The main similarity is indeed only the form of the quantum 
equations because the physical and dynamical meaning is made 
significantly different in view of the different hamiltonian 
densities (or smeared hamiltonians) respectively involved: 
in the present approach we have to do with the super-Hamiltonian, 
while in the multi-time formalism with the ADM reduced one. 

More physical insight is obtained on the
nature of the kinematical variables
by making a parallel
between our model and 
the so-called
{\em Gaussian reference fluid} \cite{KT91,I92};
we argue that the formalism below
developed corresponds to a generalization of
the Gaussian case to a {\em generic 
reference fluid} which can play the role of
a physical clock.
The physical characterization of this fluid
is provided by analysing the Hamiltonian
equations involving the kinematical variables;
as a result we show that to this generic
fluid can be associated
the energy-momentum tensor of a dust.

In Section 2 we give a review
of the quantum theory of the ``matter'' field in the 
spirit of the discussion presented in \cite{K81}, 
which constitutes 
the appropriate line of thinking for the reformulation of the 
quantum geometrodynamics developed in Section 4 and 
following a schematic derivation
of the WDE approach presented in
Section 3. 
By section 5 we compare our proposal with the 
Schr\"odinger multi-time formalism, devoting 
particular attention to a minisuperspace model. 
In Section 6 we give a discussion on the
physical nature of our time variable,
based on a generalization of the 
Gaussian fluid clock. 
Finally Section 7 is devoted to brief concluding remarks and
provides an application of the proposed quantum gravity theory 
to a very simple model.  

\section{Quantum Fields on Curved Background}  

We start by a brief review of the canonical 
quantization of a ``matter'' 
field on an assigned space-time background \cite{BD82,K81}. 

Let us consider a four-dimensional pseudo-riemannian manifold 
${\cal M}^4$, 
characterized by a metric tensor $g_{\mu \nu }(y^{\rho})$ 
($\mu , \nu , \rho = 0,1,2,3$) (having signature $-,+,,+,+$) 
and perform a $(3 + 1)$-slicing of the space-time 
\cite{ADM62,MTW73}, by a 
one-parameter family of spacelike hypersurfaces $\Sigma ^3_t$ : 
$y^{\rho } = y^{\rho }(t, x^i)$  
($i = 1,2,3 $), each of which defined by a unique  
fixed value of $t$. 
The normal field to the family of hypersurfaces 
$n^{\mu }(y^{\rho })$  
($n_{\mu }n^{\mu } = -1$)  
and the three tangent vectors to this family 
$e^{\mu }_i \equiv \partial _i y^{\mu }$ 
(we adopted the notation $\partial _i(\; ) 
\equiv \partial (\; )/\partial x^i$) 
form a reference basis in ${\cal M}^4$. 

Thus we may decompose the so-called deformation vector 
$N^{\mu } \equiv \partial _ty^{\mu }$ 
($\partial _t(\; ) \equiv  \partial (\; )/\partial t$) 
along the basis 
$\{ {\bf n }, {\bf e}_i \}$, i. e. 

\begin{equation}  
N^{\mu } \equiv \partial _ty^{\mu } = Nn^{\mu } + N^i
e^{\mu }_i  
\label{a} 
\end{equation} 

where $N(t, x^i)$ and $N^i(t, x^i)$ are respectively called the 
{\em lapse  function} and the {\em shift vector}. \\ 
By performing on the metric tensor the coordinates 
transformation 
$y^{\mu } = y^{\mu }(t, x^i)$ and using (\ref{a}) , 
the line element rewrites 

\begin{equation}  
ds^2 = g_{\mu \nu }dy^{\mu }dy^{\nu } = -N^2dt^2 + h_{ij}
(dx^i + N^i dt)(dx^j + N^jdt)  
\label{b} 
\end{equation} 

where 
$h_{ij}\equiv g_{\mu \nu }e^{\mu }_i e^{\nu }_j$ 
is the metric tensor induced on the hypersurfaces $\Sigma ^3_t$. 
By the line element (\ref{b}) we get the
expression of
the contravariant normal vector in the new coordinates
($t, x^i$),  i. e.
${\bf n} = (1/N, - N^i/N)$; 
by the inverse components of the metric
in these same coordinates, i. e.
$g^{\mu \nu }\rightarrow
\{ - 1/N^2 \; N^i/N^2\; h^{ij} - N^iN^j/N^2 \} $, 
we get the covariant vector field
${\bf n} = (- N, {\bf 0})$. 
These contravariant and covariant
expressions for the normal field
imply respectively 
$n^{\mu }e^i_{\mu } = N^i/N$ and
$n_{\mu } e^{\mu }_i = 0$

Now, within the ADM formalism,  
we analyze the quantization of a  self-interacting scalar field 
$\phi (t, x^i)$ described by a potential term $V(\phi )$ on a 
fixed gravitational background; its dynamics is summarized 
by the action 

\begin{equation}  
S^{\phi }(\pi_{\phi }, \phi ) = 
\int_{{\cal M}^4} \left\{ \pi _{\phi }\partial _t\phi  - NH^{\phi } - 
N^iH^{\phi }_i \right\}d^3xdt 
\label{c} 
\end{equation} 

where $\pi _{\phi }$ denotes the conjugate field 
to the scalar one and . 
the hamiltonian terms $H^{\phi }$ and $H^{\phi }_i$ 
read explicitly 

\begin{equation}  
H^{\phi } \equiv 
\frac{1}{2\sqrt{h}}{\pi _{\phi }}^2 + 
\frac{1}{2}\sqrt{h}h^{ij}\partial _i \phi \partial _j\phi + \sqrt{h}V(\phi )  
\,\quad H^{\phi }_i \equiv \partial _i \phi {\pi }_{\phi }  
\label{d} 
\end{equation} 

being $h\equiv det h_{ij}$. 
This action should be varied with respect ro 
$\pi _{\phi }$ and $\phi$, 
but not $N$, $N^i$ and $h^{ij}$ since the 
metric background is assigned; 
However it remains to specify the geometrical meaning 
of the lapse function and the shift vector with respect 
ro the space-time slicing; 
this aim is reached by adding to $S^{\phi }$ 
the so-called kinematical action 
\cite{K81} 

\begin{equation}    
S^k (p_{\mu }, y^{\mu }) = \int_{{\cal M}^4}\left\{   
p_{\mu }\partial _t y^{\mu } - N^{\mu }p_{\mu } \right\} d^3xdt  
\label{e} 
\end{equation} 

so getting, by (\ref{a}) the full action for the system 

\begin{equation}  
S^{\phi k}\equiv S^{\phi } + S^k = 
\int_{{\cal M}^4} \left\{ \pi _{\phi }\partial _t\phi 
+ p_{\mu }\partial _ty^{\mu } - 
N(H^{\phi } + H^k) - N^i(H^{\phi }_i + H^k_i) \right\} d^3xdt 
\label{f} 
\end{equation} 

\begin{equation}  
H^k \equiv p_{\mu }n^{\mu } \, \quad H^k_i \equiv p_{\mu }e^{\mu }_i 
\label{g} 
\end{equation} 

In the above action $n^{\mu }$ and $h^{ij}$ 
are to be regarded as assigned functionals of $y^{\mu }(t, x^i)$.
Adding the kinematical action does not affect the 
field equation for the scalar field, 
while the variations with respect to $p_{\mu }$ and $y^{\mu }$ 
provide the equation (\ref{a}) and  the evolution of the 
kinematical momentum. 

Finally, by varying, now even, with respect to $N$ and $N^i$ 
we get the Hamiltonian constraints 

\begin{equation}  
H^{\phi } = - p_{\mu }n^{\mu } \, \quad 
H^{\phi }_i = - p_{\mu }e^{\mu }_i 
\label{i} 
\end{equation} 

Clearly 
is to be assigned the following Cauchy problem  
assigned on a regular initial hypersurface $\Sigma ^3 _{t_0}$, i. e. 
$y^{\mu }(t_0 , x^i) = y^{\mu }_0(x^i)$  

\begin{equation}  
\phi (t_0,  x^i) = \phi _0(x^i) 
\, \quad 
\pi _{\phi }(t_0,  x^i) = \pi _0(x^i) 
\, \quad 
y ^{\phi }(t_0, x^i) = y ^{\mu }_0(x^i) 
\, \quad 
p_{\mu }(t_0, x^i) =  p_{\mu \; 0}(x^i) 
\label{l} 
\end{equation} 

At last, to complete the scheme of the field equations, 
we have also to specify the 
lapse function, the shift vector 
and the metric tensor 
$h_{ij} = h_{ij}(y^{\mu })$, 

This system can be easily quantized in the canonical 
formalism by assuming 
the states of the system be represented by a wave functional 
$\Psi (y^{\mu }(x^i), \phi (x^i))$ and implementing 
the canonical variables 
$\{ y^{\mu }, p_{\mu }, \phi , \pi _{\phi } \}$ to operators  
$\{ \hat{y}^{\mu } ,  \; 
\hat{p}_{\mu } = -i\hbar \delta (\; ) /\delta y^{\mu } , 
\; \hat{\phi } , \; 
\hat{\pi }_{\phi } 
= -i\hbar \delta (\; )/\delta \phi \}$. Then the quantum 
dynamics is described by the equations 

\begin{equation}   
i\hbar n^{\mu }\frac{\delta \Psi }{\delta y^{\mu }} 
= \hat{H}^{\phi } \Psi = \left[  
-\frac{\hbar ^2}{2\sqrt{h}} 
\frac{\delta  }{\delta \phi } \frac{\delta  }{\delta \phi } +  
\frac{1}{2}\sqrt{h}h^{ij}\partial _i \phi \partial _j\phi 
+ \sqrt{h}V(\phi ) \right] \Psi  \, \quad 
i\hbar e^{\mu }_i\frac{\delta \Psi }{\delta y^{\mu }} 
= \hat{H}^{\phi }_i \Psi = -i\hbar \partial _i \phi \frac{\delta \Psi }
{\delta \phi }
\label{m} 
\end{equation} 

These equations have 
$5\times \infty ^3$ degrees of freedom, corresponding 
to the values taken by the four components of $y^{\mu }$ 
and the scalar field 
$\phi$ in each point of a spatial hypersurface. 
In (\ref{m}) $y^{\mu }$ plays the role 
of ``time variable'', since it specifies  
the choice of a particular hypersurface $y^{\mu } = y^{\mu }(x^i)$. 

In view of their parabolic nature, equations (\ref{m})
have a space of solutions that, by an heuristic procedure, 
can be turned into an Hilbert space whose inner product reads 

\begin{equation}   
\langle \Psi _1 \mid \Psi _2 \rangle \equiv 
\int_{y^{\mu } = y^{\mu }(x^i)} \Psi_ 1^*\Psi _2 D\phi \, \quad 
\frac{\delta \langle \Psi _1 \mid \Psi _2 \rangle }
{\delta y^{\mu }} = 0 
\label{n} 
\end{equation} 

where $\Psi _1$ and $\Psi _2$ 
denote two generic solutions and $D\phi $ the 
Lebesgue measure defined on the $\phi$-function space.  
The above inner product induces the 
conserved functional probability 
distribution $\varrho \equiv \langle \Psi \mid \Psi \rangle$. 

The semiclassical limit of this equations (\ref{m}) 
is obtained when taking 
$\hbar \rightarrow 0$ and, by setting the wave functional as 

\begin{equation}   
\Psi = expi\left\{ \frac{1}{\hbar }\Sigma (y^{\mu }, \phi )\right\}  
\label{o} 
\end{equation} 

and then expanding $\Sigma$ in powers of  $\hbar /i$, i. e.

\begin{equation}   
\Sigma = \Sigma _0 + \frac{\hbar }{i}\Sigma _1  + 
\left( \frac{\hbar }{i}\right) ^2 \Sigma _2 + . . . 
\label{p} 
\end{equation} 

By substituting (\ref{o}) and (\ref{p}) in equations (\ref{m}), 
up to the zero-order approximation, 
we find  the Hamilton-Jacobi equations  

\begin{equation}   
- n^{\mu }\frac{\delta \Sigma _0}{\delta y^{\mu }} 
= \frac{1}{2\sqrt{h}} 
\left( \frac{\delta  \Sigma _0}{\delta \phi }\right)^2 +  
\sqrt{h}\left( \frac{1}{2}h^{ij}\partial _i \phi \partial _j\phi 
+ V(\phi ) \right)  
\, \quad 
e^{\mu }_i\frac{\delta \Sigma _0}{\delta y^{\mu }} 
= -\partial _i \phi \frac{\delta \Sigma _0}{\delta \phi }
\label{r} 
\end{equation} 

which lead to the identification $\Sigma _0 \equiv S^{\phi k}$. 

 \section{The Wheeler-DeWitt Equation} 

Now we briefly recall how the Wheeler-DeWitt approach 
\cite{D67,K81} 
faces the problem of quantizing a coupled system 
consisting of gravity and 
a real scalar field, which implies also the metric 
field now be a dynamical variable. 
The action describing this coupled system reads  

\begin{equation}   
S^{g\phi } = \int _{{\cal M}^4} \left\{ 
\pi ^{ij}\partial _th_{ij}  + \pi _{\phi }
\partial _t\phi - N(H^g + H^{\phi }) - N^i(H^g_i + H^{\phi }_i)\right\} d^3xdt  
\label{s} 
\end{equation}  

where  $\pi ^{ij}$ denotes the 
conjugate momenta to the   
three-dimensional metric tensor $h_{ij}$ and the gravitational 
super-hamiltonian 
$H^g$ and super-momentum $H^g_i$ takes the explicit form 

\begin{equation}   
H^g \equiv \frac{16\pi G}{c^3}G_{ijkl}\pi^{ij}\pi^{kl} - \frac{c^3}{16\pi G}
\sqrt{h}{}^3R  \, \quad G_{ijkl} \equiv \frac{1}{2\sqrt{h}}
(h_{ik}h_{jl} + h_{il}h_{jk} - h_{ij}h_{k})   
\label{t} 
\end{equation}  

\begin{equation}   
H^g_i \equiv -2{}^3\nabla _j\pi ^j_i 
\label{u} 
\end{equation}  

In the above expressions ${}^3R$ and ${}^3\nabla (\; )$ 
denote respectively 
the Ricci scalar and the covariant derivative constructed 
by the 3-metric $h_{ij}$, while 
$G$ is the Newton constant,

Since now $N$ and $N^i$ are, in principle, 
dynamical variables, they have to be varied, so leading 
to the constraints 
$H^g + H^{\phi } = 0$ and  $H^g_i + H^{\phi }_i = 0$ 
which are equivalent to the   
$\mu -0$-components of the Einstein equations 
and therefore play the 
role of constraints for the Cauchy data. 
It is just this restriction 
on the initial values problem, a peculiar 
difference between the previous case, 
at fixed background, and the present 
one; in fact, now, on the regular hypersurface $t = t_0$, 
the initial conditions 
$\{ \phi _0(x^i), \;  \pi _0(x^i), \;  
h_{ij\; 0}(x^i), \; \pi _{ij\; 0}(x^i)\}$ 
can not be assigned freely, 
but they must verify on $\Sigma ^3_{t_0}$ the four relations 
$\{ H^g + H^{\phi }\} _{t_0} = \{ H^g_i + H^{\phi }_i \}_{t_0} = 0$.  

Indeed behaving like Lagrange multipliers, 
the lapse function and the shift vector 
have not a real dynamics and their specification 
corresponds to assign 
a particular slicing of ${\cal M}^4$, 
i. e. a system of reference. 

In order to quantize this system 
we assume that its states be represented  
by a wave functional $\Psi (\{ h_{ij}\}, \phi )$ 
(the notation $\{ h_{ij}\}$ means all the 3-geometries 
connected by a  3-diffeomorphism ) 
and implement the 
canonical variables to operators acting on 
this wave functional 
(in particular we set $h_{ij}\rightarrow \hat{h}_{ij}, \; 
\pi ^{ij} \rightarrow \hat{\pi }^{ij}\equiv 
-i\hbar \delta (\; )/\delta h_{ij}$).\\ 
The quantum dynamics of the system is then induced by imposing 
the operatorial translation of the classical constraints, 
which leads respectively to the Wheeler-DeWitt equation 
and to the supermomentum constraint one \cite{D67}:  

\begin{equation}   
(\hat{H}^g + \hat{H}^{\phi })\Psi = 0 \, \quad 
(\hat{H}^g_i + \hat{H}^{\phi }_i)\Psi = 0 
\label{v} 
\end{equation}  

which to be explicited requires a specific choice for the 
{\em normal ordering} of the operators.  

Due to its hyperbolic nature this formulation 
of the quantum dynamics has some limiting feature \cite{D97} , 
which we summarize by the following  
four points:\\ 
i) It doesn't exist any general  procedure  allowing 
to turn the space 
of the solutions into an Hilbert one 
and so any appropriate general notion of  
functional probability distribution is prevented.\\  
ii) The WDE doesn't contain any dependence 
on the variable $t$ or on the function $y^{\mu }$, 
so loosing its evolutive character along the 
slicing $\Sigma ^3_t$. Moreover individualizing 
an internal variable which 
can play the role of ``time '' is an ambiguous procedure which 
doesn't lead to a general prescription.\\  
iii) The natural semiclassical limit, 
provided by a wave functional of the form   

\begin{equation}   
\Psi (\{ h_{ij} \}, \phi ) = 
expi\left\{ \frac{\Sigma (\{ h_{ij}\} . \phi )}{\hbar }\right\} 
\label{w} 
\end{equation}  

leads, up to the zero order of approximation 
$\Sigma \equiv \Sigma _0(\{ h_{ij}\} , \phi )$, 
to the right Hamilton-jacobi equation 
for the classical scalar tensor action $S^{g\phi }$,  
but, in the limit corresponding to a fixed background  
$N=N^*\; N^i= N^{i*}\; h_{ij} = h^*_{ij}$, 
the WDE doesn't provide neither the Hamilton-Jacobi 
equation (\ref{r}) 
expected associated to the semiclassical limit 
$\Psi (\{h^*_{ij}\} , \phi )
= expi\{(\Sigma _0(\phi ))/\hbar \}$ ($\hbar \rightarrow 0$), 
neither the appropriate quantum dynamics on a fixed 
background (\ref{m}). in correspondence to the wave functional 
$\Psi = \Psi (\{h^*_{ij}\},\phi ) \equiv \chi (\phi )$, 
which instead provides  

\begin{equation} 
\hat{H}^{\phi }\chi = 0 \, \quad
\hat{H}^{\phi }_i\chi = 0 
\label{y} 
\end{equation}  

This discrepancy is due to the absence of a dependence on $y^{\mu }$  
in the action $S^{g\phi }$, 
which can not be clearly restored anyway.\\ 
iv)At last we stress what is to be regarded as an 
intrinsic inconsistency of the approach 
above presented:
the WDE is based on the primitive notion 
of space-like hypersurfaces,  
i. e. of a time-like normal field, 
which is in clear contradiction with 
the random behavior of a quantum metric field \cite{ZJ96}; 
indeed the space 
or time character of a vector 
becomes a precise notions only in the limit of 
a perturbative quantum gravity theory.  
This remarkable ambiguity leads us 
to infer that there is  inconsistency 
between  the requirement of a wave equation 
(i. e. a wave functional) 
invariant, like the WDE one, 
under space diffeomorphisms and time displacements 
on one hand, and, on the other one, the 
(3 + 1)-slicing representation of the global manifold.     

The existence of these shortcomings in the WDE 
approach, induces us to search for 
a better reformulation of the quantization 
procedure which addresses 
the solution of the above indicated  four points 
as prescriptions to write down new dynamical quantum constraints. 

\section{Reformulation of the Quantum Geometrodynamics} 

We start by observing that, within the framework of a functional 
approach, a covariant quantization of the 4-metric field is 
equivalent to take the wave amplitude 
$\Psi = \Psi(g_{\mu \nu }(x^{\rho }))$; 
in the WDE approach, by adopting the ADM slicing 
of the space-time, the problem is restated in terms of the 
following replacement

\begin{equation}
\Psi (g_{\mu \nu }(x^{\rho })) \rightarrow 
\Psi (N(t, x^l), N^{i}(t, x^l), h_{ij}(t, x^l)) 
\, \quad . 
\label{sss} 
\end{equation} 

Then, since the lapse function $N$ and the shift 
vector $N^i$ are cyclic variables, i. e. 
their conjugate momenta $p_N$ and $p_{N^i}$ vanish identically, 
we get, on a quantum level, the following restrictions: 

\begin{equation}
p_N = 0, \; p_{N^i} = 0 \, \quad \Rightarrow \, \quad 
\frac{\delta \Psi }{\delta N} = 0, \; 
\frac{\delta \Psi }{\delta N^i} = 0  
\, \quad ; 
\label{sss1} 
\end{equation} 

by other words, the wave functional $\Psi$ 
should be independent of 
$N$ and $N^i$. Finally, the super-momentum constraint leads 
to the dependence of $\Psi$ on the 3-geometries $\{ h_{ij} \}$ 
(instead on a single 3-metric tensor $h_{ij}$).\\ 
The criticism to the WDE approach, developed at the point iv) of 
the previous section, concerns with 
the ill-defined nature of the 
replacement (\ref{sss}). The content of this section is 
entirely devoted to reformulate the quantum geometrodynamics, 
by preserving the (3 + 1)-representation of the space-time, 
but avoiding the ambiguity above outlined in the WDE approach.  

Since, as well known \cite{K81}, the Hamiltonian constraints   
$H^g + H^{\phi } = 0$ and  $H^g_i + H^{\phi }_i = 0$,  
once satisfied on the initial hypersurface, are 
preserved  by the remaining Hamilton equations, 
then the above variational principle (\ref{s}) 
would be equivalent to a new one with assigned 
lapse function and shift vector 
(so loosing the Hamiltonian constraint), 
but with the specification of a Cauchy problem 
satisfying these constraints. 
Therefore   
we face the question of introducing an external 
temporal dependence  
in the quantum dynamics of gravity, 
by addressing, in close analogy to the 
approach discussed in Section $2$, 
the kinematical action (\ref{e}) 
as till present even in the geometrodynamics framework. 
Therefore we adopt already on a classical level the full action 

\begin{equation}   
S^{g\phi k} = \int _{{\cal M}^4} \left\{ 
\pi ^{ij}\partial _th_{ij}  + \pi _{\phi }   
\partial _t\phi + p_{\mu }\partial _ty^{\mu } -  
N(H^g + H^{\phi } + p_{\mu }n^{\mu }) - N^i(H^g_i + H^{\phi }_i + 
p_{\mu }e^{\mu }_i)\right\} d^3xdt  
\label{s99} 
\end{equation}  

where $n^{\mu}$ is to be regarded as a functional of 
$y^{\mu }(t, x^i)$, but, differently from the case of  
a fixed background, now $h_{ij}$ (like $\phi$) are dynamical 
variables taken on $t$ and $x^i$. 
The variation of this action with respect to the variables 
$h_{jj}\; \pi_{ij} \;  \phi \; \pi _{\phi }$ 
provides the standard 
Einstein-scalar dynamics, the 
variation with respect to $y^{\mu }$ and 
$p_{\mu }$ gives rise respectively to the relations 
(\ref{a}) and 
$\partial _t p_{\mu } = 
-p_{\nu }N\delta n^{\nu }/\delta y^{\mu }
+ \partial _i(N^ip_{\mu })$  and
finally we have to vary with respect
to $N$ and $N^i$,
so getting the constraints 

\begin{equation}  
H^g + H^{\phi } = - p_{\mu }n^{\mu } \, \quad 
H^g_i + H^{\phi }_i = - p_{\mu }e^{\mu }_i 
\label{z} 
\end{equation} 

However we observe that the equation
$\partial _t p_{\mu } = 
-p_{\nu }N\delta n^{\nu }/\delta y^{\mu }
+ \partial _i(N^ip_{\mu })$
is a linear and homogeneous first order
differential system which 
ensures that if $p_{\mu }$ is chosen
to be zero on the
initial hypersurface 
$\Sigma ^3_{t_0}$  
then it will remain zero during
the whole evolution
\footnote{We remark that if $p_{\mu }$
vanishes on the initial
hypersurface, then this equation implies that
all its initial time derivatives vanish.}
and the considered theory reduces 
to ordinary general relativity. 
Thus we see how, as argued  above, the difference in addressing 
the presence of the kinematical action, 
respectively to the canonical case 
relies only on a different structure of the Cauchy problem.\\ 

The key point in our reformulation of the canonical 
quantization of gravity consists  
of assuming (\ref{z}) be the appropriate 
quantum constraints and individualizing the 
restriction to be imposed on the initial wave functional to get 
$p_{\mu } \equiv 0$ 
when constructing its semiclassical expansion. 

In agreement with these considerations 
we require that the quantum states 
of the system be represented by a wave functional 
$\Psi = \Psi (y^{\mu },\; h_{ij}, \; \phi )$ 
which, once performed the 
canonical variables translation into the corresponding operators, 
satisfy the super-hamiltonian and 
super-momentum constraints 

\begin{equation}   
i\hbar n^{\mu }\frac{\delta \Psi }{\delta y^{\mu }} 
= (\hat{H}^g + \hat{H}^{\phi }_i)\Psi 
\, \quad 
i\hbar e^{\mu }_i\frac{\delta \Psi }{\delta y^{\mu }} 
= (\hat{H}^g_i + \hat{H}^{\phi }_i)\Psi  
\label{a1} 
\end{equation} 

The second of the above equations exprimes the 
non-invariance of the wave functional under space 
diffeomorphisms (i. e. $x^{i\prime } = x^{i\prime }(x^i)$)
and therefore, since we expect the quantization procedure, 
even in a $(3 + 1)$-representation of space-time,   
should yet preserve the tensorial nature of the 3-geometry, then 
it is natural to require the left-hand-side of 
of this equation vanish 
(on a classical level it corresponds to the restriction 
$p_{\mu }\propto n_{\mu }\; \Rightarrow \; 
p_{\mu }e^{\mu }_i = 0$); 
thus we take, as describing 
the quantum geometrodynamics, the following system of  
functional differential equations: 

\begin{equation}   
i\hbar n^{\mu }\frac{\delta \Psi }{\delta y^{\mu }} 
= (\hat{H}^g + \hat{H}^{\phi }_i)\Psi 
\, \quad 
(\hat{H}^g_i + \hat{H}^{\phi }_i)\Psi = 0 
\, \quad \Psi = \Psi (\{ h_{ij}\} , y^{\mu }) 
\label{a1x} 
\end{equation} 

where now the wave functional is taken again on the 
3-geometries ($\{ h_{ij}\}$) related by the 3-diffeomorphisms.\\
These ($4\times \infty ^3$) 
equations, which correspond to a natural extension of the 
Wheeler-DeWitt approach, have the 
fundamental feature 
that the first of them is now a parabolic one 
and it is just this their property which allows to overcome 
some of the above discussed limits of the WDE.

Though this set of equations provides a satisfactory 
description of the 3-geometries quantum dynamics, 
nevertheless it turns out convenient and physically meaningful 
to take, by (\ref{a}), the wave functional evolution along 
a one-parameter family of spatial hypersurfaces 
filling the universe. 

By (\ref{a}) the firsts of equations (\ref{a1x}) can be 
rewritten as follow 
(taking into account even the second one of (\ref{a1x})):

\begin{equation}   
i\hbar \frac{\delta \Psi }{\delta y^{\mu }}\partial _ty^{\mu } 
= N(\hat{H}^g + \hat{H}^{\phi })\Psi 
\label{b1} 
\end{equation} 

Now this set of equations can be (heuristically) rewritten 
as a single one by 
integrating over the hypersurfaces $\Sigma ^3_t$, i. e. 

\begin{equation}   
i\hbar \partial _t\Psi = 
i\hbar \int _{\Sigma _t^3} \left\{ 
\frac{\delta \Psi }{\delta y^{\mu }}\partial _ty^{\mu } 
\right\} d^3x =    
\hat{{\cal H}}\Psi  \equiv 
\left[ \int _{\Sigma _t^3} 
N(\hat{H}^g + \hat{H}^{\phi }) d^3x \right] \Psi 
\label{c1} 
\end{equation} 

The above equations (\ref{c1}) and (\ref{a1x}) 
show  how in the present approach the 
wave functional is no longer invariant under 
infinitesimal displacements of the time variable.  

Let us now show that the operator  
$\hat{{\cal H}}$ is an hermitian one, i. e., 
in the bra-ket Dirac notation, it verifies the relations 

\begin{equation}   
\langle \Psi _1\mid \hat{{\cal H}}\Psi _2\rangle = 
\langle \hat{{\cal H}}\Psi _1\mid \Psi _2\rangle 
\label{e1} 
\end{equation} 

being $\Psi _1$ and $\Psi _2$ two generic solutions of 
equations (\ref{c1}) and (\ref{a1x}). 

We start by choosing in $\hat{H}^g$ 
the following normal ordering for its 
kinetic part 

\begin{equation}   
G_{ijkl} \pi ^{ij}\pi ^{kl} \rightarrow 
- \hbar ^2\frac{\delta \; }{\delta h_{ij}}\left( G_{ijkl}
\frac{\delta (\; )}{\delta  h_{kl}}\right) 
\label{f1} 
\end{equation} 

Hence we have 

\begin{equation}   
\langle \Psi _1\mid \hat{{\cal H}}\Psi _2\rangle \equiv 
\int_{{\cal F}_t}DhD\phi \int_{\Sigma ^3_t}d^3x\Psi _1^*N
(\hat{H}^g + \hat{H}^{\phi })\Psi _2
\label{g1} 
\end{equation} 

where ${\cal F}_t$ denotes the functional space
($\{ h_{ij}\}, \phi $), as referred 
to the hypersurface $\Sigma ^3_t$,     
$DhD\phi$ the associated Lebesgue measure and $\Psi ^*$ 
the complex 
conjugate wave functional, which satisfies the hermitian 
conjugate equation of (\ref{c1}) and (\ref{a1x}) ones.  

By observing that, 
as shown in \cite{K81} and easily checkable by repeating 
the analysis below performed for the gravitational terms, 
$\hat{H}^{\phi }$ is an hermitian operator, 
then it remains to be analyzed the term 
(the functional integrals on $\phi$ and $h$ 
commute with each other and both commute with the 
space integral )
\footnote{The analysis of this section and of the following one 
has an heuristic character 
due to its functional approach and it should 
be made rigorous by an appropriate discretization on a lattice.}:

\begin{equation}   
\int_{{\cal F}_t}Dh \left\{ 
\int_{\Sigma ^3_t}\Psi_1^*  
\left[ -\frac{16\pi G\hbar ^2}{c^3}
\frac{\delta \; }{\delta h_{ij}}
\left( G_{ijkl}\frac{\delta \Psi _2}{\delta h_{kl}}\right) - 
\frac{\sqrt{h}c^3{}^3R}{16\pi G} \Psi _2\right] 
\right\} 
\label{h1} 
\end{equation} 

Since $G_{ijkl} = G_{klij}$, it is easy to check the relation 

\begin{equation}   
\int_{\Sigma ^3_t} d^3x\left[ 
\Psi _1^*\frac{\delta \; }{\delta h_{ij}}
\left( G_{ijkl}\frac{\delta \Psi _2}{\delta h_{kl}}\right) 
\right] = 
\label{i1} 
\end{equation} 

\begin{equation} 
\int_{\Sigma ^3_t} d^3x\left[ 
\frac{\delta \; }{\delta h_{ij}}\left( 
\Psi _1^*G_{ijkl}\frac{\delta \Psi _2}{\delta h_{kl}} - 
\Psi _2G_{ijkl}\frac{\delta \Psi _1^*}{\delta h_{kl}} \right) + 
\frac{\delta \; }{\delta h_{ij}}
\left( G_{ijkl}\frac{\delta \Psi _1^*}{\delta h_{kl}}\right) 
\Psi _2 \right] 
\label{i1x} 
\end{equation} 

and then, assuming 
(as done for the matter field $\phi$ \cite{K81}) 
the validity in the 
functional space ($\{ h_{ij} \}$) of the functional Gauss theorem  
(the wave functional is required to vanish on the ``boundary'  
of this space): 

\begin{equation}   
\int_{{\cal F}_t}Dh \int_{\Sigma ^3_t}d^3x
\frac{\delta \;}{\delta h_{ij}}\left( ...\right) = 0 
\label{l1} 
\end{equation} 

we conclude the proof that $\hat{{\cal H}}$ 
is an hermitian operator. 

On the base of this results,  
by (\ref{c1}) we get 

\begin{equation}   
\partial _t \langle \Psi_1 \mid \Psi_2\rangle 
\equiv 
\int_{\Sigma^3_t}d^3x\partial _ty^{\mu } 
\frac{\delta \; }{\delta y^{\mu }}
\langle \Psi_1 \mid \Psi_2\rangle = 
\label{o1} 
\end{equation} 

\begin{equation}   
\langle \partial _t\Psi _1\mid \Psi _2\rangle + 
\langle \Psi _1\mid \partial _t\Psi _2\rangle = 
\frac{i}{\hbar }(\langle \hat{{\cal H}}\Psi _1\mid 
\Psi _2\rangle - \langle \Psi _1\mid 
\hat{{\cal H}}\Psi _2\rangle ) = 0 
\label{o11} 
\end{equation} 

and the generic character of the deformation vector allows us to 
write the fundamental conservation law 

\begin{equation}  
\frac{\delta \langle \Psi _1 \mid \Psi_2\rangle }
{\delta y^{\mu }} = 0 
\label{o99} 
\end{equation} 

Thus we defined an inner product which turns  
the space of the solutions to equations (\ref{c1}) 
and (\ref{a1x}) 
into an Hilbert space (so removing the shortcoming i) of the WDE). 
Indeed now it is possible to define the notion of a 
conserved functional probability 
distribution as: 

\begin{equation}  
\varrho (y^{\mu }, \{ h_{ij} \}, \phi ) \equiv 
\Psi ^*\Psi \, \quad 
\langle \Psi \mid \Psi \rangle = 1 \, \quad 
\frac{\delta \langle Psi  \mid \Psi \rangle }
{\delta y^{\mu }} = 0 
\label{p1} 
\end{equation} 

Let us now reformulate the dynamics described by the equations 
(\ref{c1}) and (\ref{a1x}) by means of the eigenvalues 
problems for the equations (\ref{a1x}) 
To this end we expand $\Psi $ in the following functional 
representation: 

\begin{equation}  
\Psi (y^{\mu }, \{ h_{ij}\} , \phi ) = 
\int_{ {}^*{\cal Y}_t}D\omega \Theta (\omega ) 
\chi _{\omega }(\{ h_{ij}\} , \phi )
exp\left\{  
\frac{i}{\hbar }\int _{\Sigma^3_t}d^3x
\int dy^{\mu }(\omega n_{\mu })\right\}  
\label{q1} 
\end{equation} 

where $D\omega$ denotes the Lebesgue measure in the 
functional space ${}^*y_t$ of the conjugate function  
$\omega (x^i)$, $\Theta$ a functional valued in this domain 
and $n_{\mu }$ denotes the 
covariant normal vector; 
indeed, once assigned $n^{\mu }(y^{\mu }) $, 
the field $n_{\mu }$ can be written, in general, only formally 
in a quantum space-time. 

Substituting this expansion into equations (\ref{a1x}) 
we get the eigenvalues problems

\begin{equation}  
(\hat{H}^g + \hat{H^{\phi }})\chi _{\omega } 
= \omega \chi _{\omega } \, \quad 
(\hat{H}^g_i + \hat{H}^{\phi }_i)\chi _{\omega } = 0 
\label{r1} 
\end{equation} 

Here $\omega (x^i)$ is not a 3-scalar, but it 
transforms, under 3-diffeomorphisms, like $\hat{H}^g$ or 
$\hat{H}^{\phi }$, so ensuring that $\omega d^3x$, 
as it should, be an invariant quantity.\\ 
Now we observe that, by (\ref{a}),
equation (\ref{q1}) rewrites

\begin{equation}  
\Psi (y^{\mu }, \{ h_{ij}\} , \phi ) = 
\int_{ {}^*{\cal Y}_t}D\omega 
\Theta (\omega )\chi _{\omega }(\{ h_{ij}\} , \phi )
exp\left\{  
\frac{i}{\hbar }\int _{\Sigma^3_t}d^3x
\int _{t_0}^{t}dt^{\prime }
\partial _{t^{\prime }}y^{\mu }(\omega n_{\mu })\right\} =
\label{t1} 
\end{equation} 

\begin{equation}  
\Psi (\{ h_{ij}\} , \phi , t) = 
\int_{ {}^*{\cal Y}_t}D\omega \Theta (\omega )\chi _{\omega }
(\{h_{ij}\} , \phi )
exp\left\{  
-\frac{i}{\hbar }\int _{t_0}^t dt^{\prime }
\int _{\Sigma^3_t}d^3x
(N\omega )\right\}   
\label{u1} 
\end{equation} 

being $t_0$ an assigned initial ``instant.`\\ 
To the same result we could arrive by choosing, without any loss 
of generality, the coordinates system ($t, x^i$), i. e. 
$y^0\equiv t \; , y^i\equiv x^i$; 
indeed, for this system, the spatial hypersurfaces 
have equation $t = const,$ i. e. 
$dy^{\mu }\rightarrow (dt, 0, 0, 0)$
and we have $n_0 = N$.
By other words the wave functional (\ref{u1}) 
is to be interpreted directly in terms of the time variable $t$, i. e. 
$\Psi (\{ h_{ij}\} , \phi , t)$ and, 
in fact, it turns out solution of the wave equation 

\begin{equation}  
i\hbar \partial _t \Psi (\{ h_{ij}\} , \phi , t) = \hat{{\cal H}}
\Psi(\{ h_{ij}\} , \phi , t )
\label{v1} 
\end{equation} 

The expansion (\ref{u1}) of the wave functional and  the 
eigenvalues problems (\ref{r1}) completely describe  
the quantum dynamics of the 3-geometries. 

Now to conclude our analysis, it needs to 

recognize which restriction
should be imposed on the initial wave functional, say 
$\Psi (\{ h_{ij}\} , \phi \} , t_0) =  
\Psi _0 (\{ h_{ij}\} , \phi )$ 
($t = t_0$ defining the initial hypersurface) 
to get in the classical limit the ordinary general relativity, 
i. e. the quantum counterpart of the condition $p_{\mu 0} = 0$. 

To this aim we preliminary observe that, being $\hat{{\cal H}}$ 
an hermitian operator 
the same remains valid for $\hat{H}^g$ (by identical proof) and 
the functionals $\chi_{\omega }$ 
are expected to be an orthonormal basis, i. e. 
$\langle \chi _{\omega }\mid \chi _{\omega ^{\prime }}\rangle 
= \Delta (\omega - \omega ^{\prime })$  
($\Delta$ denoting the Dirac functional), 
on which we can expand $\Psi _0$; 
in fact we get the functional relation 
$\Theta (\omega ) = \langle \Psi _0\mid \chi _{\omega }\rangle$. 

As next step 
we express the wave functional as follows

\begin{equation}  
\Psi \equiv \sqrt{\rho }exp\left\{ i\frac{\sigma }{\hbar }
\right\} 
\label{w1} 
\end{equation} 

being $\rho$ and $\sigma$ respectively 
the modulus and the real phase (up to $\hbar $) of $\Psi$. 
By substituting (\ref{w1}) into the equation (\ref{v1}) 
we get for its real part an expression of the form: 

\begin{equation}  
-\partial _t \sigma =  
\int _{\Sigma ^3_t}\left\{ 
N\hat{HJ}\sigma + 
\left[ \frac{N}{\sqrt{\rho }}(\hat{H}^g + \hat{H}^{\phi })  
\right] \sqrt{\rho }\right\} d^3x  
\label{x1} 
\end{equation} 

where by $\hat{HJ}$ we denote 
the Hamilton-Jacobi operator;
we stress how, in the above equation,  
the terms relative to the functional $\rho$ 
contain, differently from the Hamilton-Jacobi one, 
$\hbar$, as well as the imaginary 
part of (\ref{v1}) 
(which is an evolutive equation for $\rho$). \\ 
Now we observe that once assigned 
$\sigma _0 \equiv \sigma (\{ h_{ij}\} , \phi , t_0)$, 
equation (\ref{x1}) allows to calculate on 
$\Sigma^3_{t_0}$ 
$(\partial _t\sigma )\mid _{t = t_o}$ 
and, by iteration, all higher order time derivatives;
thus we can expand $\sigma$ in powers of $t$ 
near enough to the initial hypersurface, i. e. : 

\begin{equation}  
\sigma = \sum_{n=0}^{\infty }\frac{1}{n!}
(\partial _t)^n \sigma \mid _{t=t_0}(t - t_0)^n 
\label{z1} 
\end{equation} 

This expression permits to extend the solution toward the 
future  by reassigning the Cauchy data in $t_0 + \Delta t 
\; , \Delta t\ll 1$, and then iterating the procedure  
indefinitely. But if we require that $\sigma _0$ satisfies 
the restriction 

\begin{equation}  
\hat{HJ}\sigma _0 = 0 
\label{a3} 
\end{equation}  

than we get 

\begin{equation}  
\partial _t \sigma \mid _{t=t_0} = terms \sim O(\hbar ) + ... 
\, \quad \Rightarrow \, \quad 
\sigma = \sigma _0 + terms\sim O(\hbar )(t - t_0) + ... 
\label{b3} 
\end{equation}  

This result ensures in the semiclassical limit 
$\hbar \rightarrow 0$ the envolutive 
nature of the wave functional 
phase, i. e. $\sigma = \sigma _0$ 
(the same is not true for the modulus $\rho$
which remains an evolutive variable even in this limit). 
Thus an initial wave functional whose modulus is a generic one, 
but whose phase satisfies 
the Hamilton-Jacobi equation (\ref{a3}) 
provides a quantum evolution 
compatible with the classical limit of general relativity.\\ 
However we have to note that to retain the invariance 
under the 3-diffeomorphisms of the quantization procedure 
prevents, in general, the achievement of a correct limit 
for the quantum field theory on a fixed background; 
indeed (\ref{a1x}) provides, on a fixed background, the right 
dynamics (\ref{m}) only in those reference frames for which 
$N^i = 0$.  

We conclude our reformulation of the 
quantum geometrodynamics, by emphasizing how, 
a restriction on the initial wave functional phase $\sigma _0$ 
(as the one required to get the classical 
limit of general relativity),
does not correspond to a real loss of physical 
degrees of freedom on $\Sigma^3_{t_0}$, 
since the only meaningful 
information we can provide on the initial 
quantum configuration of 
the system, consists of the functional probability distribution 
$\rho _0$ (which, to get the classical 
limit, should be peaked around a 
specific solution of the Einstein equations). 

\section{The Multi-time and Schr\"odinger Approach} 

In this section we provide a schematic formulation of the 
so-called multi-time approach 
and of its smeared Schr"odinger version, 
in view of a comparison with the proposal of previous section. 

The multi-time formalism is based on 
the idea that many gravitational 
degrees of freedom appearing in the classical geometrodynamics 
have to be not quantized because 
are not real physical ones; 
indeed we have to do with 10 $\times \infty ^3$ 
variables, i. e. the values of the functions 
$N\, N^i\, h_{ij}$ in each point of the hypersurface $\Sigma ^3$, 
but it is well-known that the gravitational field 
possesses only $4\time \infty ^3$ physical degrees of freedom 
(in fact the gravitational waves have,
in each point of the space, only two independent
polarizations and satisfy second order equations).\\ 
The first step is therefore to extract the real  
canonical variables by the transformation 

\begin{equation}
\left\{ h_{ij}\, \pi ^{ij} \right\} \, \rightarrow \, 
\left\{ \xi ^{\mu }\, \pi _{\mu } \right\} \quad  
\left\{ H_r \, P ^r  \right\} \quad  
\mu = 0,1,2,3 \, r = 1,2 
\, \quad , 
\label{xax1}
\end{equation}

where $H_r , P^r$ are the four real degrees of freedom, while 
$\xi ^{\mu }\, \pi _{\mu }$ play the role of embedding variables. 

In terms of this new set of canonical 
variables, the gravity-``matter'' action (\ref{s})
rewrites

\begin{equation}   
S^{g\phi } = \int _{{\cal M}^4} \left\{ 
\pi _{\mu }\partial _t\xi ^{\mu } + 
P ^r\partial _tH _r +   
\pi _{\phi }\partial _t\phi - N(H^g + H^{\phi }) - 
N^i(H^g_i + H^{\phi }_i)\right\} d^3xdt  
\, \quad , 
\label{xax2}
\end{equation}  

where
$H^g = H^g ({\xi }^{\mu }, {\pi }_{\mu },
H_r, P^r)$.
and 
$H^g_i = H^g_i({\xi }^{\mu }, {\pi }_{\mu },
H_r, P^r)$.

Now we provide an ADM reduction of the
dynamical problem by solving
the Hamiltonian constraint for the momenta
$\pi _{\mu }$

\begin{equation}   
{\pi }_{\mu } + h _{\mu }({\xi }^{\mu }, 
H_r, P^r, \phi , \pi _{\phi }) = 0                      
\label{xax3}
\end{equation}  

Hence the above action takes the reduced form

\begin{equation}   
S^{g\phi } = \int _{{\cal M}^4} \left\{ 
P ^r\partial _tH _r +   
\pi _{\phi }\partial _t\phi - 
h _{\mu }\partial _t\xi ^{\mu } \right\}d^3xdt 
\, \quad . 
\label{xax4}
\end{equation}  

Finally the lapse function and the shift vector
are fixed by the Hamiltonian equations lost with the
ADM reduction, as soon as, the functions
$\partial _t {\xi }^{\mu }$ are assigned.
A choice of particular relevance is to set
$\partial _t {\xi }^{\mu } = \delta ^{\mu }_0$
which leads to

\begin{equation}   
S^{g\phi } = \int _{{\cal M}^4} \left\{ 
P ^r\partial _tH _r +   
\pi _{\phi }\partial _t\phi - 
h _{0}\right\}d^3xdt 
\, \quad . 
\label{xax5}
\end{equation}  

The canonical quantization of the model
follows by replacing all the Poisson brackets with 
the corresponding commutators;
if we assume that the states
of the quantum system are represented
by a wave functional
$\Psi = \Psi ({\xi }^{\mu }, H_r, \phi )$,
then the evolution is described by the equations

\begin{equation}   
i\hbar \frac{\delta \Psi }{\delta {\xi }^{\mu }} =
\hat{h}_{\mu } \Psi
\, \quad , 
\label{xax6}
\end{equation}  

where $\hat{h}_{\mu }$ are the operator version
of the classical Hamiltonian densities.\\
In its smeared formulation the multi-time
approach reduces to the following Schr\"odinger
equation

\begin{equation}   
i\hbar \partial _t\Psi =
\hat{{\cal h}}\Psi \, \quad
\Psi = \Psi (t, H_r, \phi ) 
\, \quad . 
\label{xax7}
\end{equation}  

Here $\hat{{\cal h}}$ denote the quantum
correspondence to the smeared hamiltonian

\begin{equation}   
{\cal h} = \int _{{\cal M}^4} \left\{ 
h _{\mu }\partial _t\xi ^{\mu } \right\}d^3xdt 
\, \quad . 
\label{xax8}
\end{equation}  

Now, observing that the first of equations
(\ref{a1x}) can be rewritten as follows 

\begin{equation}
i\hbar \frac{\delta \Psi }{\delta y^{\mu }} 
= -n_{\mu }(\hat{H}^g + \hat{H}^{\phi }_i)\Psi 
\, \quad , 
\label{xax9} 
\end{equation} 

it exists a correspondence between the above
multi-time approach and our proposal, 
viewed by identifying the formulas
(\ref{s99})-(\ref{xax5}),
(\ref{xax9})-(\ref{xax6}) and 
(\ref{c1})-(\ref{xax7}).
But the following two key differences appear
evident:
i) the embedding variables $y^{\mu }$
are added by hand, while the corresponding ones 
${\xi }^{\mu }$ 
come from non-physical degrees of freedom;
ii) the hamiltonians ${\cal H}$ and ${\cal h}$
(as well as their corresponding densities)
describe very different dynamical situations.

We show explicitly the parallel between these two
approaches by their implementation
in a minisuperspace model: a Bianchi type
IX Universe
containing a self-interacting scalar field. 
By using Misner variables
($\alpha $, $\beta _+$, $\beta _-$) \cite{MTW73} 
the classical action describing this system reads:

\begin{equation}
S = \int \left\{
p_{\alpha }\dot{\alpha } +
p_{\beta _+}\dot{\beta _+} +
p_{\beta _-}\dot{\beta _-} +
p_{\phi }\dot{\phi } - cNe^{-3\alpha }
\left(
-p_{\alpha }^2 +
p_{\beta _+}^2 +
p_{\beta _-}^2 +
p_{\phi }^2 + V(\alpha , \beta _{\pm }, \phi )
\right)
\right\} dt 
\, \quad c = const.
\, \quad ,
\label{xax10} 
\end{equation} 

where $\dot{(\; )}\equiv d(\; )/dt$ 
and the precise form of the potential term $V$
is not relevant for our discussion.\\
For this model, since the Hamiltonian density is
independent of the spatial coordinates, then the
multi-time approach and its
smeared Schr\"odinger version overlap, the same
being true in our formalism. 

In the spirit of our proposal
the quantum dynamic of this model
is described by the equation 

\begin{equation}
i\hbar \partial _t \Psi = cNe^{-3\alpha }
\hbar ^2\left\{
\partial _{\alpha }^2 -
\partial _{\beta _+}^2 -
\partial _{\beta _- }^2 - 
\partial _{\phi }^2 + V\right\} \Psi 
\, \quad
\Psi = \Psi (t, \alpha , \beta _{\pm }, \phi ) 
\, \quad ,
\label{xax11} 
\end{equation} 

to which it should be added the
restriction that the initial wave function phase
$\sigma _0 = \sigma _0(\alpha , \beta _{\pm },
\phi )$
satisfies the Hamilton-Jacobi equation

\begin{equation}
\left\{
-(\partial _{\alpha })^2 +
(\partial _{\beta _+})^2 +
(\partial _{\beta _- })^2 + 
(\partial _{\phi })^2 \right\} \sigma _0 + V = 0
\, \quad . 
\label{xax12} 
\end{equation} 

In this scheme $N(t)$ is
an arbitrary function of the label
time to be specified when fixing a reference. 

To set up the multi-time approach
we have to preliminarily perform
an ADM reduction of the dynamics (\ref{xax10}).
By solving the Hamiltonian constraint obtained
varying $N$, we find the relation

\begin{equation}
-p_{\alpha } \equiv h_{ADM} =
\sqrt{p_{\beta _+}^2
+ p_{\beta _-}^2 + p_{\phi }^2 + V}
\, \quad .
\label{xax13} 
\end{equation} 

Therefore  action (\ref{xax10}) rewrites as 

\begin{equation}
S = \int \left\{
p_{\beta _+}\dot{\beta _+} +
p_{\beta _-}\dot{\beta _-} +
p_{\phi }\dot{\phi } - \dot{\alpha }
h_{ADM}\right\} dt 
\, \quad ,
\label{xax14} 
\end{equation} 

Thus we see how $\alpha $ plies the role of
an embedding variable (indeed it is related to the
Universe volume), while $\beta _{\pm }$
are the real gravitational degrees of freedom
(they describe the Universe anisotropy).\\
By one of the Hamiltonian equation lost in the ADM
reduction (i. e. when varying $p_{\alpha }$ in
(\ref{xax10})), we get

\begin{equation}
\dot{\alpha } = -2cNe^{-3\alpha }p_{\alpha } =
2cNe^{-3\alpha }h_{ADM} 
\, \quad . 
\label{xax15} 
\end{equation} 

Hence by setting $\dot{\alpha } = 1$,
we fix the lapse function as

\begin{equation}
N =
\frac{e^{3\alpha }}{2ch_{ADM}} 
\, \quad . 
\label{xax16} 
\end{equation} 

The quantum dynamics in the multi-time approach is
summarized by the equation

\begin{equation}
i\hbar \partial _{\alpha }\Psi =
\sqrt{ - \hbar ^2(\partial _{\beta _+}^2 +
\partial _{\beta _-}^2 +
\partial _{\phi }^2 ) + V}\Psi
\, \quad \Psi = \Psi (\alpha , \beta _{\pm },
\phi ) 
\, \quad .
\label{xax17} 
\end{equation} 

We stress that in this multi-time
approach the variable
$\alpha $, i. e. the volume of the Universe,
behaves as a ``time``-coordinate
and therefore the quantum dynamics can not avoid
the Universe reaches 
the cosmological singularity
($\alpha \rightarrow - \infty $). On the other hand, 
in the formalism we proposed,
$\alpha $ is on the same footing of the other
variables and are admissible ``stationary states``
for which it is distributed in probabilistic way.\\
This feature reflects a more general
and fundamental difference
existing between the two approaches:
the multi-time formalism violates
the geometrical nature of the
gravitational field in view of real physical 
degrees of freedom, while
the proposed quantum dynamics
implements this idea only up to the lapse function
and the shift vector, but 
preserves the
geometrical origin of the 3-metric field.

\section{Physical Interpretation of the Model}

A fundamental question we have to answer
is about the physical meaning of the kinematical
variables adopted in the present reformulation
of the quantum dynamics.
Indeed in the multi-time approach the
corresponding embedding variables have no
physical meaning since they do not
represent any physical degree
of freedom, but simply
equivalent ways to represent the same
real dynamics in the space-time.\\
However, in our case, these variables have 
been added by hand to the geometrodynamics
and provide, in general, a violation of the
classical symmetries;
more precisely, by restricting, as 
in (\ref{a3}),
the initial phase of the wave functional,
we violate the symmetry of general relativity
only on a quantum level and they are
restored on a classical limit,
when $\hbar \rightarrow 0$. 

A viable point of view is to
regard this feature of the dynamics as a
prescription of the nature;
by other words we may argue that the
only experimental knowledge at
our disposition concerns the classical dynamics
only (which therefore is to be restored anyway),
but noting we know, in principle, 
about the quantum behavior of gravity
(which, in the proposed paradigm we
prescribe to be summarized
by the present equations).

Another, more physical, point of view relies on
observing that any dynamical term able to deform    
the dynamics of an assigned system with        
different issues for its evolution,
should have a precise physical meaning and
can be identified with some kind of field.
In what follows, we address
this last way of thinking
and argue the kinematical term can be
interpreted as a matter fluid, in close
analogy with the so-called
{\em Gaussian reference fluid} which was
first proposed by K. Kuchar and his
collaborators, see \cite{KT91,I92}.
In this spirit the restriction (\ref{a3})
is no longer a key requirement for the
consistency of the theory, but the validity of
the idea requires that this material component be
experimentally detected. 

We start by schematically reviewing
the Gaussian reference approach.
In a system of Gaussian coordinates
$\{ T\, , X^i\} $ ($i=1,2,3$),
the line element of a generic gravitational field
takes the form

\begin{equation}
ds^2 = - dT^2 + h_{ij}(T, X^k)dX^idX^j
\, \quad ,
\label{xbx1}
\end{equation}

and therefore such a system is obtained
from a generic one by imposing
the covariant Gaussian constraints 

\begin{equation}
g^{\mu \nu}\partial _{\mu }T
\partial _{\nu }T = - 1 
\, \quad  
g^{\mu \nu}\partial _{\mu }T
\partial _{\nu }X^i = 0 
\, \quad .
\label{xbx2}
\end{equation}

In the above formula we adopted the notation
$\partial _{\mu }\equiv \partial \; /
\partial y^{\mu }$, being $y^{\mu }$
generic coordinates to which
it corresponds the metric tensor
$g_{\mu \nu }$.

The idea now consists into add this
constraint to the gravity-``matter'' action
(i. e. into requiring the latter 
be restricted to Gaussian references) 
via four Lagrangian multipliers $M$ and $M^i$;
thus we consider an additional
action term of the form

\begin{equation}
S^{(G)} = \int _{{\cal M}^4}\left\{ 
M(g^{\mu \nu}\partial _{\mu }T
\partial _{\nu }T - 1) + 
M^i(g^{\mu \nu}\partial _{\mu }T
\partial _{\nu }X^i) \right\}d^3xdt 
\, \quad .
\label{xbx3}
\end{equation}

A careful analysis of the whole variational
problem, leads \cite{KT91} to dynamical
constraints of the form

\begin{equation}
p_T + h_T = 0 
\, \quad 
p_{X^i} + h_{x^i} = 0 
\, \quad ,
\label{xbx4}
\end{equation}

where 
$p_T$ and $p_{X^i}$ denote the conjugate momenta 
respectively of the variables $T$ and $X^i$,
while $h_T$ and $h_{X^i}$ correspond
to linear combinations of the super-hamiltonian
and the super-momentum with coefficients
depending on $T$, $X^i$ and the 3-metric
$h_{ij}$.
The linearity of the above  constraints
with respect to the added momenta, allows to
perform a satisfactory
quantization of the model in the spirit
of a Schr\"odinger-like formalism.\\
The physical interpretation of this proposal 
relies on regarding the added term as a
kind of fluid which interact with gravity.
Though this point of view is affected
by some shortcomings
(indeed this fluid has a quite peculiar dynamics),
nevertheless it provides
an appropriate notion of reference fluid clock.

The link of our formalism with the above
Gaussian reference fluid consists of
the observation that in
Gaussian coordinates the spatial Hypersurfaces have
equation
$T(y^{\mu }) = const.$ and therefore
the quantity 
$\partial _{\mu }T$
describes the covariant normal field $n_{\mu }$,
while the gradients
$\partial _{\mu }X^i$
are provided via the reciprocal
vectors $e_{\mu }^i$ as
$\partial _{\mu }X^i =
e^i_{\mu } + (N^i/N)n_{\mu }$.
Hence it is natural to argue that
our kinematical term corresponds to
a generic reference fluid, having an
associated time-like vector 
$n^{\mu }$, for which the first of
(\ref{xbx2}) is generalized to the
normalization condition
$g^{\mu \nu }n_{\mu }n_{\nu } = - 1$;
the second condition (\ref{xbx2})
does not hold for a generic fluid since we have
$g^{\mu \nu }n_{\mu }
\partial _{\mu }X^i = 
n^{\mu }[e_{\mu }^i + (N^i/N)n_{\mu }] =
-N^i/N \neq 0$.
Our parallel with the Gaussian reference fluid
becomes precise by stressing how,
in our case, 
the normalization condition for $n^{\mu }$
should not be added as a constraint
(by some Lagrangian multiplier);
indeed it is ensured by the relation (\ref{a}), obtained
when varying the kinematical action with
respect to $p_{\mu }$.\\
We recall that the dynamics of the
{\em generic fluid reference} is described by the
Hamiltonian equations 

\begin{equation}  
\partial _ty^{\mu } = Nn^{\mu } + N^i
\partial _iy^{\mu }
\, \quad
\partial _t p_{\mu } =
-p_{\nu }N\delta n^{\nu }/\delta y^{\mu }      
+ \partial _i(N^ip_{\mu }) 
\, \quad .
\label{xbxc}
\end{equation} 

Once assigned the vector $n^{\mu }$
as a functional of $y^{\mu }$ and the
functions $N(t, x^i)$ $N^i(t, x^i)$,
we can solve the first of these
equations to get the
kinematical variables  
$y^{\mu } = y^{\mu }(t, x^i)$;
Hence by substituting
this information into the second Hamiltonian
equation and solving it, we get the momentum 
$p_{\mu } = p_{\mu }(t, x^i)$.

In order to outline the relation
existing between this system of Hamiltonian
equations and the dynamics of a fluid,
we multiply the second of (\ref{xbxc}) by
$n^{\mu }$, so getting via the first one 

\begin{equation}  
n^{\mu }[\partial _t p_{\mu } +
p_{\nu }N\delta n^{\nu }/\delta y^{\mu }      
- \partial _i(N^ip_{\mu })] =
\partial _t(p_{\mu }n^{\mu }) -
\partial _i(N^ip_{\mu }n^{\mu }) = 0 
\, \quad . 
\label{xbxd}
\end{equation} 

Hence by setting
$p_{\mu }n^{\mu } = - \bar{\omega }$ 
and labeling with barred indices $\bar{\mu},
\bar{\nu},...$ all the quantities
in the coordinates $t, x^i$, we rewrite the
above equation as follows

\begin{equation}  
\partial _t(p_{\mu }n^{\mu }) -
\partial _i(N^ip_{\mu }n^{\mu }) =
-\partial _{\bar{\mu }}(N\bar{\omega }
n^{\bar{\mu }}) = 0 \;
\Rightarrow \nabla _{\bar{\mu }}
(\varepsilon n^{\bar{\mu }}) = 0 
\, \quad , 
\label{xbxe}
\end{equation} 

where
$\varepsilon \equiv \bar{\omega }/\sqrt{h}$
denotes a real 3-scalar and
$\nabla $ refers to the covariant
4-derivative.
By the covariance of this equation,
we should have in general

\begin{equation}
\nabla _{\mu }
(\varepsilon n^{\mu }) = 
-n^{\mu }\nabla _{\nu }t_{\mu }^{\nu } = 0
\, \quad , 
\label{xbxf}
\end{equation} 

being $t_{\mu }^{\nu }\equiv
\varepsilon
n_{\mu }n^{\nu }$ the 
energy-momentum tensor of a dust
with energy density $\varepsilon$ and
four-velocity $n^{\mu } = n^{\mu }(t, x^i)$.
The real correspondence of this tensor
with the kinematical term, i. e.
with the {\em generic fluid reference},
comes out when observing, first that the requirement
$e^{\mu }_i\delta \Psi/\delta y^{\mu } = 0$
implies, on a classical level
$p_{\mu }e^{\mu }_i = 0$ and then how 
the kinematical term $-p_{\mu }n^{\mu }$,
appearing on the right-hand-side of the
super-Hamiltonian constraint
can be rewritten in the expressive form

\begin{equation}
-p_{\mu }n^{\mu } = \bar{\omega } =
\sqrt{h}\varepsilon =
\sqrt{h}t_{\mu \nu }n^{\mu }n^{\nu } =
\frac{1}{N^2}
\sqrt{h}t_{\mu \nu }
\partial _ty^{\mu }\partial _ty^{\nu } = 
\frac{1}{N^2}t_{\bar{0}\bar{0}}
= -\sqrt{h}t_{\bar{0}}^{\bar{0}}
\label{xbxg}
\end{equation} 

(of course, apart from 
$t_{\bar{0}\bar{0}}$, all the other components
$t_{\bar{\mu }\bar{\nu }}$ vanish identically). 

Therefore the super-Hamiltonian constraint
acquires the familiar expression 

\begin{equation}
H^g + H^{\phi } - \bar{\omega } = 
H^g + H^{\phi } + 
\sqrt{h}t_{\bar{0}}^{\bar{0}} = 0 
\, \quad .
\label{xbxh}
\end{equation} 

Thus we see how the {\em generic reference fluid},
having the energy-momentum tensor of a dust fluid,
contributes to the full super-Hamiltonian by
a quantity like the one due 
to a space-time
dependent cosmological term, coinciding with 
its energy density. 

In the light of these considerations, 
the requirement that $p_{\mu }$ vanishes
identically implies that
$\varepsilon \equiv 0$ and therefore 
such a reference fluid does not
interact with gravity by its ``energy-momentum''; 
by a purely phenomenological point of view,
in this case,
it behaves like a test ``matter`` field,
whose kinematics is fixed by (\ref{a}). 

This identification with a reference fluid
allows us to achieve the fundamental result of 
upgrading the formal time
$y^{\mu }$ to a real physical clock. 

Finally we stress the following two points:\\ 
i) The momentum equation is equivalent
to the conservation law for the dust
energy-momentum tensor and therefore provides the
behavior of $\varepsilon $.
For a Gaussian system this equation yields
$\varepsilon = c(x^i)/\sqrt{h}\; \Rightarrow \;
\bar{\omega }\equiv c(x^i)$,
being $c$ a generic space function;
This behavior is just
the one of a dust energy density,
but it is worth noting that
our analysis does not oblige $\varepsilon$
to be a positively defined quantity.\\
ii) The classical limit of
the super-Hamiltonian equation (\ref{a1x})
provides an Hamilton-Jacobi equation where
the eigenvalue $\omega $ coincides
(see (\ref{xbxh}) with the
function $\bar{\omega }$ taken on a specific
hypersurface. 

\section{Concluding Remarks and a Simple model}

As outcoming of our analysis, we get a reformulation 
of the canonical quantization of gravity in 
which is removed the so-called 
``frozen formalism'' typical of the WDE, i. e. 
the wave functional becomes evolutive along a 
one-parameter family 
of spatial hypersurfaces filling the space-time and can 
be expressed in terms of a direct dependence on the parameter $t$ 
labeling the slicing (this result provides a solution to the 
shortcoming of the WDE emphasized at the point ii)); 
indeed the existence of an Hilbert space associated 
to the solutions of the restated equation can be regarded 
as a consequence of the non-frozen formalism here obtained 
(resulting into a parabolic nature of the super-hamiltonian 
quantum constraint).\\ 
Instead of these successful results, the WDE shortcoming 
indicated at the point iii) is overcome only with respect to 
those reference frames where the shift vector vanishes; 
the reasons for this incompleteness are to be regarded as a 
direct consequence of retaining in the quantum geometrodynamics 
the invariance of the wave functional under 
the 3-diffeomorphisms.\\ 
Of course the relaxation of this restriction is 
a possible issue of an extended theory; 
here we do not address this point because it implies 
some non-trivial complications in constructing an Hilbert 
space (at least on our heuristic level), but overall 
because it would correspond to leave {\em a priori} the notion 
of a covariant 3-geometrodynamics, without any strong 
physical motivation to pursue this way 
(indeed we are forced to relax the time displacement invariance 
by the incompatibility of a quantum space-time and the 
(3 + 1)-slicing notion). 

However the main goal of our analysis is achieved by removing 
the shortcoming of the WDE stated at the point iv), 
i. e. now the 
quantization procedure takes place in a fixed system 
(indeed only the lapse function should be specified, 
while the shift 
vector can assume a generic value) and no 
ambiguity survives about 
the time-like character of the normal field; 
by other words, in this new approach it is possible to 
quantize the 3-geometry field on a fixed family of 
spatial hypersurfaces (corresponding to its 
evolution in the space-time), 
because this quantization scheme does not contradicts the 
strong assumption of a (3 + 1)-slicing 
of the 4-dimensional manifold. 

The discussion presented in Section 6, 
about the interpretation of the kinematical
variables as a generic reference fluid clock, 
has the very important merit to
transform a working formalism into
a possible experimental issue; 
however it should be supported by 
further investigation on the 
physical consequences the clock fluid  has 
when referred to specific contexts:
indeed the question about the appropriate definition
of a reference frame 
in a quantum space-time is a really subtle one 
(see for instance \cite{R91a,R91b}); 

In spite of this available physical issue 
(ensured by the fluid interpretation), 
we emphasize how to have found the (non-physical)
restriction on the initial wave functional phase (\ref{a3}), which ensures 
the classical limit coinciding with general relativity, 
is essential for the consistency of the whole approach; 
in fact the physical meaning of this result concerns the 
fundamental achievement of restoring, on a classical level, 
the invariance of the theory under the time displacements, 
which is instead broken by the quantum dynamics. 

We conclude by observing how the functional nature of all our 
approach implies it has 
(like in the WDE) 
a heuristic value; but it appear 
rather  reasonable that it can be made rigorous when 
reformulated, in a discrete approach, as a theory on 
a suitable ``lattice''.\\ 
Though is out of the aim of this paper to face this problem,
(to be regarded as a fundamental subject of further 
investigations), 
nevertheless we here suggest that the best method to 
reformulate the quantum dynamics on a discrete level, seems to 
be via the Regge calculus \cite{R61,R97} as applied to the 
3-geometry field. 
The physical justification for a discrete approach to 
quantum gravity relies on the expectation that the 
space-time has a {\it lattice structure} 
(or a granular morphology) on a Planckian scale. 

At the end of this work we provide an application of 
the obtained theory to the quantization of a very simple model, 
described by the following line element:

\begin{equation} 
ds^2 = N(t)^2dt^2 - r(t)^{4/3}\delta _{ij}dx^idx^j 
\, \quad , 
\label{ttt}
\end{equation}

where $\delta _{ij}$ denote the Kr\"onecker matrix and 
the flat hypersurfaces $t = const.$ are taken to have a 
closed topology, i. e. $0\le x^i < 2\pi L$ ($i = 1,2,3$) 
($L = const.$ is the radius of the three cyrcles and has 
the dimensionality of a length). 
We allow $r$ belongs to the positive real axis 
$0\le r < \infty $ (since the metric is invariant 
under the exchange $r\rightarrow -r$. \\ 

The ADM action describing the vacuum dynamics of this 
``1-dimensional'' model, reads in the simple form 

\begin{equation} 
S = -\frac{\pi ^2L^3c^3}{2G}\int 
\dot{r}^2dt = 
\int \left\{ p_r\dot{r} - \frac{GN}{2\pi ^2L^3c^3}(-p_r^2)
\right\}dt \, \quad ,
\label{ttt1}
\end{equation}

where $\dot{(\; )}\equiv \partial _t (\; )$ and $p_r$ 
denotes the conjugate momentum to the variable $r$.\\ 
From a classical point of view this model corresponds to 
the Euclidean 3-space (to which is associated a generic 
time variable) since 
$\dot{r} \propto p_r = 0 \Rightarrow r = r_0 = const.$  and 
the corresponding Hamilton-Jacobi equation and solution read 

\begin{equation} 
\left( \frac{d\sigma }{dr}\right) ^2 = 0 \Rightarrow 
\sigma = \sigma _0 = const, 
\, \quad . 
\label{ttt2}
\end{equation}

The quantum dynamics of the model is described by equation 
(\ref{c1}), which, in the present case reduces to the 
very simple form:

\begin{equation} 
i\hbar \partial _t \Psi (t,r) = 
\frac{N\hbar^2}{2\mu }\partial _r^2\Psi (t, r) \, \quad 
\mu \equiv  \frac{\pi ^2 L^3c^3}{G} 
\, \quad . 
\label{ttt3}
\end{equation}

Apart from the negative nature of its hamiltonian, we see how 
this quantum gravity model corresponds to the 
free nonrelativistic 
particle; the general solution of 
the above equation reads in the 
following wave-packet form:

\begin{equation} 
\Psi (t, r) = \frac{1}{\sqrt{\hbar}} 
\int_{-\infty }^{\infty }dp \varphi (p) exp\left\{ 
\frac{i}{\hbar }\left( pr + \frac{p^2}{2\mu }
\int_{t_0}^{t}N(t^{\prime })dt^{\prime }
\right) \right\} 
\, \quad , 
\label{ttt4}
\end{equation}

where $\varphi (p)$ is determined by the relation 

\begin{equation} 
\varphi (p) =  \frac{1}{\sqrt{h}}
\int_{0}^{\infty }dr 
\Psi _0(r)exp\left\{ -i\frac{pr}{\hbar }\right\} 
\, \quad \Psi _0(r)\equiv \Psi (t=0, r) 
\, \quad . 
\label{ttt5}
\end{equation}

Since $\Psi$ (as well as $\Psi _0$) 
should verify the boundary conditions 
$\Psi (r=0) = 0$ (in addition to 
$\Psi (r\rightarrow \infty) = 0$),  
then $\varphi$ should be antisymmetric in its argument, i. e. 
$\varphi (-p) = -\varphi (p)$.\\ 

In terms of the expression (\ref{ttt5}) and by choosing 
the synchronous gauge $N = c$ ($t\rightarrow T$), 
we may rewrite the wave 
function (\ref{ttt4}) as follows 
(taking into account the expression for $\mu $ and 
setting $T_0 = 0$): 

\begin{equation} 
\Psi (T, r) = e^{i\sigma _0/\hbar} 
\frac{1}{2\pi } 
\int_{0}^{\infty }dr^{\prime }\chi _0(r^{\prime })
\int_{-\infty }^{\infty }dx exp\left\{ 
i\left[ x(r - r^{\prime }) + x^2
\frac{cT{l_{Pl}}^2}{2\pi^2L^3}
\right] \right\} 
\, \quad , 
\label{ttt6}
\end{equation}

where $l_{Pl}$ denotes the Planck length  
($l_{Pl} \equiv \sqrt{G\hbar /c^3}$) 
and, in agreement with the restriction (\ref{a3}), we set 

\begin{equation} 
\Psi _0 = \chi _0e^{i\sigma _0/\hbar} 
\, \quad , 
\label{ttt7}
\end{equation}

being $\chi _0$ a real function subjected 
only to the boundary conditions 
$\chi _0(r=0) = 0$ and $\chi _0(r\rightarrow \infty ) = 0$. 

At last, it is worth noting that, for any fixed $T$, 
takes place the (intriguing) limit 

\begin{equation} 
\lim _{(cT{l_{Pl}}^2/L^3)\rightarrow 0}\Psi(T, r) = \Psi _0(r) 
\, \quad , 
\label{ttt8}
\end{equation}

where the initial probability distribution $\chi _0^2$ 
can eventually be 
regarded as strongly peaked around a fixed value $r = r_0$. 

\vspace{2cm}

We are very grateful to Bryce DeWitt for his valuable 
comment on the topic faced by this paper.

\end{document}